\documentclass[12pt]{article}
\usepackage{epsfig}
\usepackage[hang,bf,small]{caption}
\DeclareGraphicsExtensions{.eps.gz,.eps,.ps,.ps.gz}
\parskip=\itemsep               

\newcommand{\beq}{\begin{equation}}
\newcommand{\eeq}{\end{equation}}
\newcommand{\beqar}[1]{\begin{eqnarray}\label{#1}}
\newcommand{\eeqar}{\end{eqnarray}}

\newcommand{\si}{\sigma}

\newcommand{\as}{\alpha_S}

\def\eq#1{{Eq.~(\ref{#1})}}

\def\arnps#1#2#3{  {\it Ann. Rev. Nucl. Part. Sci. }{\bf #1} (19#2) #3}
\def\npb#1#2#3{    {\it Nucl. Phys. }{\bf B#1} (19#2) #3}

\def\prd#1#2#3{    {\it Phys. Rev. }{\bf D#1} (19#2) #3}

\def\zpc#1#2#3{    {\it Z. Phys. }{\bf C#1} (19#2) #3}

\relax
\begin{document}
\title{
{\Large \bf   Diffractive Dissociation   }\\\
{ \Large \bf  from Non-Linear Evolution  in DIS on Nuclei}}
\author{
{\large  ~ E.~Levin\thanks{e-mail:
leving@post.tau.ac.il}~~$\mathbf{{}^{a),\,b)}}$ \,~\,and\,\,~
M. ~Lublinsky\thanks{e-mail: mal@post.tau.ac.il ~~mal@tx.technion.ac.il}
~~$\mathbf{{}^{a)}}$}\\[4.5ex]
{\it ${}^{a)}$ HEP Department}\\
{\it  School of Physics and Astronomy}\\
{\it Raymond and Beverly Sackler Faculty of Exact Science}\\
{\it Tel Aviv University, Tel Aviv, 69978, ISRAEL}\\[2.5ex]
{\it ${}^{b)}$ DESY Theory Group}\\
{\it 22603, Hamburg, GERMANY}\\[4.5ex]
}

\maketitle
\thispagestyle{empty}

\begin{abstract}
A process of single diffractive dissociation off nuclei is
considered on a basis of solutions to the nonlinear evolution
equation. The relevant saturation scales $Q_{s\,A}^D(x,x_0)$ are
determined and their dependences on Bjorken $x$, atomic number
$A$, and minimal rapidity gap $Y_0\equiv \ln 1/x_0$ are
investigated. The solutions are shown to possess a geometrical
scaling in a broad kinematic region for $x$ well below $x_0$.

The ratio $\sigma^{diff}/\sigma^{tot}$ is computed for several
nuclei. We predict that at $x\simeq 10^{-4}$ this ratio is of the
order 25\% which is much larger compared to the one of proton.
This result indicates a possibility to observe a very strong
nuclear shadowing.
 \end{abstract}
\thispagestyle{empty}
\begin{flushright}
\vspace{-19.5cm}
TAUP 2710-02\\
\end{flushright}
\newpage
\setcounter{page}{1}

\section{Introduction}
\setcounter{equation}{0}

During the last years there has been a significant growth in the
interest to a new phase of QCD associated with high parton density
\cite{HDQCD,asy}. This interest is mostly related to availability
of new low $x$ $e-p$ DIS data. Another source of information on
QCD dynamics at high parton density is due to nuclei which can
provide high density effects at comparatively lower energies. The
researches on nuclear shadowing has been recently accelerated due
to the start of the RHIC collider.

In the present paper we concentrate on a process of single
diffraction dissociation off nuclei which plays an important role
in revealing QCD dynamics at high parton density. Diffractive
inclusive production in DIS is believed to be very sensitive  to
shadowing effects \cite{KM} and, in fact, is a measure of these
effects (due to the AGK cutting rules \cite{AGK}). We are going to
investigate the cross section of diffractive dissociation at fixed
impact parameter and as a final result compute the ratio of the
total inclusive diffraction to the total inclusive cross section
in DIS off nuclei.

The total deep inelastic cross section  is related to the dipole
cross section

\beq \label{F2} \si(x,Q^2)\,\,\,=\,\,\int\,\,d^2 r_{\perp} \int
\,d z\,\, |\Psi^{\gamma^*}(Q^2; r_{\perp}, z)|^2 \,\,\sigma_{\rm
dipole}(r_{\perp}, x)\,, \eeq where the QED wave functions
$\Psi^{\gamma^*}$ of the virtual photon are well known
\cite{MU94,DOF3,WF}. The  dipole cross section  is given by the
integral over the impact parameter $b$:

\beq \label{TOTCX} \sigma_{\rm dipole}(r_{\perp},x)
\,\,=\,\,2\,\,\int\,d^2 b\,\,N(r_{\perp},x;b)\,, \eeq

where $N$ stands for imaginary part of the  dipole-target
interaction amplitude for dipole of the size $r_\perp$ scattered
elastically  at the impact parameter $b$. This function is a
subject to a nonlinear quantum evolution equation first derived by
Balitsky and Kovchegov (BK) \cite{BA,KO}. The BK equation was
studied both asymptotically \cite{asy} and numerically
\cite{num,LGLM}. For the purposes of the present paper we will use
a numerical solution of this equation obtained in Ref. \cite{LL}.

The total cross section of single diffractive dissociation is
similarly defined:

\beq \label{F2D} \si^{diff}(x,x_0,Q^2)\,\,\,=\,\,\int\,\,d^2
r_{\perp} \int \,d z\,\, |\Psi^{\gamma^*}(Q^2; r_{\perp}, z)|^2
\,\,\sigma_{\rm dipole}^{diff}(r_{\perp}, x,x_0)\,, \eeq

and

\beq \label{DDCX} \sigma_{\rm dipole}^{diff}(r_{\perp},x,x_0)
\,\,=\,\,2\,\,\int\,d^2 b\,\,N^D(r_{\perp},x,x_0;b)\,. \eeq

The function $N^D$ is a partial cross section for the
dipole-target diffractive scattering with the minimal rapidity gap
$Y_0=\ln 1/x_0$. A nonlinear quantum evolution equation for $N^D$
was derived in Ref. \cite{KL} and then rederived in Ref.
\cite{Kovner}:

\begin{eqnarray}
 &    N^D({\mathbf{x_{01}}},Y,Y_0;b)  \,=\,  N^2({\mathbf{x_{01}}},Y_0;b)\,
{\rm e}^{-\frac{4
\,C_F\,\as}{\pi} \,\ln\left( \frac{{\mathbf{x_{01}}}}{\rho}\right)(Y-Y_0)}\,
+\frac{C_F\,\as}{\pi^2}\,\int_{Y_0}^Y dy \,  {\rm e}^{-\frac{4
\,C_F\,\as}{\pi} \,\ln\left( \frac{{\mathbf{x_{01}}}}{\rho}\right)(Y-y)} \nonumber \\
&  \nonumber \\
&\times
 \int_{\rho} \, d^2 {\mathbf{x_{2}}}
\frac{{\mathbf{x^2_{01}}}}{{\mathbf{x^2_{02}}}\,
{\mathbf{x^2_{12}}}}
[\,2\,  N^D({\mathbf{x_{02}}},y,Y_0;{ \mathbf{ b-
\frac{1}{2}
x_{12}}})
+  N^D({\mathbf{x_{02}}},y,Y_0;{ \mathbf{ b - \frac{1}{2}
x_{12}}})  N^D({\mathbf{x_{12}}},y,Y_0;{ \mathbf{ b- \frac{1}{2}
x_{02}}}) \nonumber \\ & \nonumber \\
&- 4 \, N^D({\mathbf{x_{02}}},y,Y_0;{ \mathbf{ b - \frac{1}{2}
x_{12}}})  N({\mathbf{x_{12}}},y;{ \mathbf{ b- \frac{1}{2}
x_{02}}})+2\, N({\mathbf{x_{02}}},y;{ \mathbf{ b -
\frac{1}{2}
x_{12}}})  N({\mathbf{x_{12}}},y;{ \mathbf{ b- \frac{1}{2}
x_{02}}})
]\,. \nonumber \\ \label{DDEQ}
\end{eqnarray}

The lhs. of Eq. (\ref{DDEQ}) is a partial cross section of the
diffractive dissociation for dipole of the size $r_\perp\equiv
\mathbf{x_{01}}\equiv |\mathbf{x_{0}}-\mathbf{x_{1}}|$ and
rapidity $Y\equiv \ln 1/x$. The rhs. of Eq. (\ref{DDEQ}) describes
quantum evolution in which the original dipole first splits to two
dipoles and then the latter  scatter off the target. The
ultraviolet cutoff $\rho$ is defined to regularize the integral,
but it does not appear in physical observables. The following
numerical results are checked to be indpendent on a choice of $\rho$.

When the rapidity fills the whole rapidity gap ($Y=Y_0$),
diffraction is reduced to  pure elastic interaction

\beq\label{DDini} N^D(r_\perp,x_0,x_0;b)\,=\,N^2(r_\perp,x_0;b)\,.
\eeq

Eq. (\ref{DDini}) serves as initial condition to the evolution
(\ref{DDEQ}).

For a proton target a numerical solution of Eq. (\ref{DDEQ}) was
found and investigated for the first time in Ref. \cite{DDLL}. In
the following paper \cite{DDLL1} the ratio
$\sigma^{diff}/\sigma^{tot}$ was computed and in a certain
kinematic domain shown to be energy independent in agreement with
the HERA data \cite{HERA}. In the present paper we report on the
numerical solution of Eq. (\ref{DDEQ}) for nucleus targets and
repeat the program of Refs. \cite{DDLL, DDLL1}.

The paper is organized as follows. In the next Section (2) we
present solutions of Eq. (\ref{DDEQ}) for several nuclei. Section
3 is devoted to the determination of the saturation scale and its
properties. Geometrical scaling is studied in Section 4. The ratio
$\sigma^{diff}/\sigma^{tot}$ is computed in Section 5. Section 6
present a theoretical discussion of  the saturation scale.
 The last Section (7) is concluding.

\section{Solution of the non-linear equation}

In this section we report on the numerical solution of Eq.
(\ref{DDEQ})    for several  nuclear targets. We consider six real
nuclei: $Ne_{20}$, $Ca_{40}$, $Zn_{70}$, $Mo_{100}$, $Nd_{150}$,
and $Au_{197}$. All the details about nuclear profile functions
are borrowed from Ref. \cite{NDATA} and are summarized in our
paper \cite{LL}. As in a series of our previous papers, solutions
to Eq. (\ref{DDEQ}) are obtained by the method of iterations
proposed in Ref. \cite{LGLM}. The constant value for the strong
coupling constant $\as=0.25$ is always used. For the function $N$,
which is an input in Eq. (\ref{DDEQ}), we use a solution of the BK
equation obtained in Ref. \cite{LL}. The solutions are computed
for $10^{-4} \le x_0\le 10^{-2}$ and within the kinematic region
$10^{-7}\le x \le x_0$ and transverse distances up to a few fermi.

The function $N^D_A$\footnote{The subscript $A$ indicates a
relation to a nuclear target with atomic number $A$.} is formally
a function of four variables: the energy gap $x_0$, the energy
variable $x$, the transverse distance $r_\perp$, and the impact
parameter $b$. In order to simplify the problem we will proceed
similarly to the treatment of the $b$-dependence of the function
$N_A$ \cite{LL} and $N^D$ for proton \cite{DDLL}. Namely, we use
the ansatz which preserves the very same $b$-dependence as
introduced  in the initial conditions: \beq \label{NDb}
 N^D_A(r_\perp,x,x_0; b)\,=\, (1\,-\,e^{-\kappa^D_A(x,x_0,r_\perp)\, S_A(b)/S_A(0)})^2,
\eeq with \beq \label{kappaD}
\kappa^D_A(x,x_0,r_\perp)\,=\,-\,\ln(1\,-\,\sqrt{\tilde
N^D_A(r_\perp,x,x_0)}). \eeq

The function $\tilde N^D_A(r_\perp,x,x_0)$ is a solution of Eq.
(\ref{DDEQ}) but with no dependence on the forth variable. The
initial conditions for $\tilde N^D_A(r_\perp,x,x_0)$ are set at
$b=0$.

\begin{figure}[htbp]
\begin{tabular}{c c c c}
 \epsfig{file=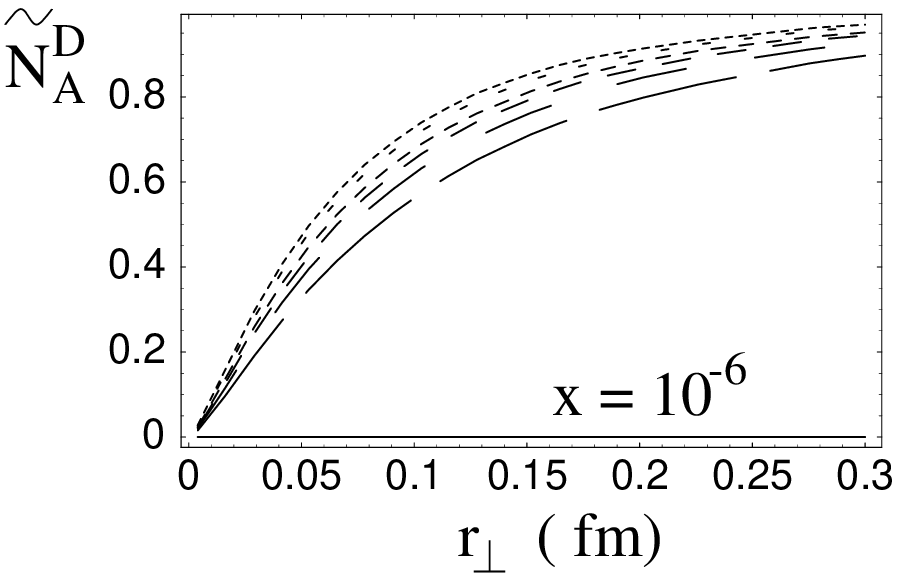,width=40mm, height=42mm}&
\epsfig{file=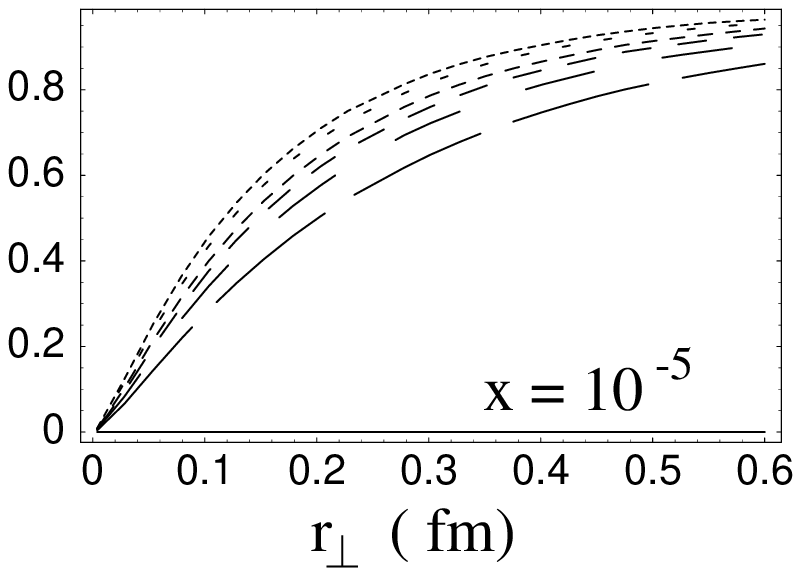,width=40mm, height=42mm}&
 \epsfig{file=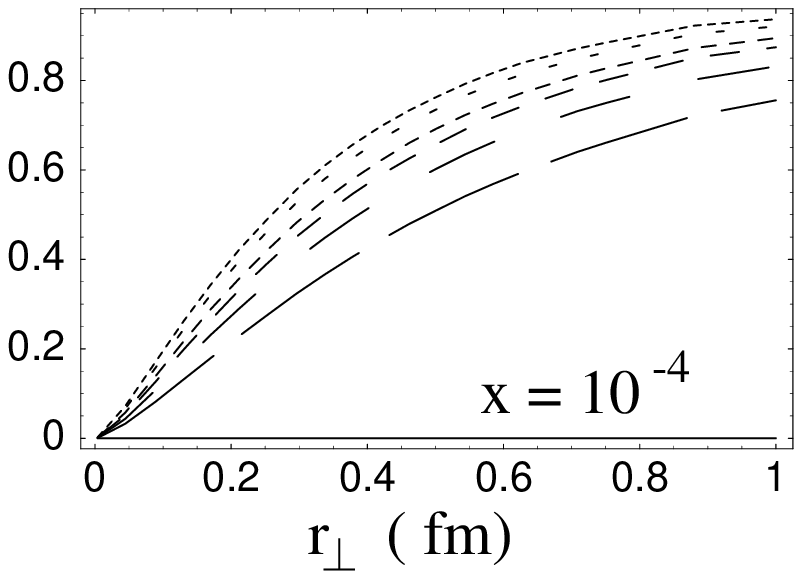,width=40mm, height=42mm} &
\epsfig{file=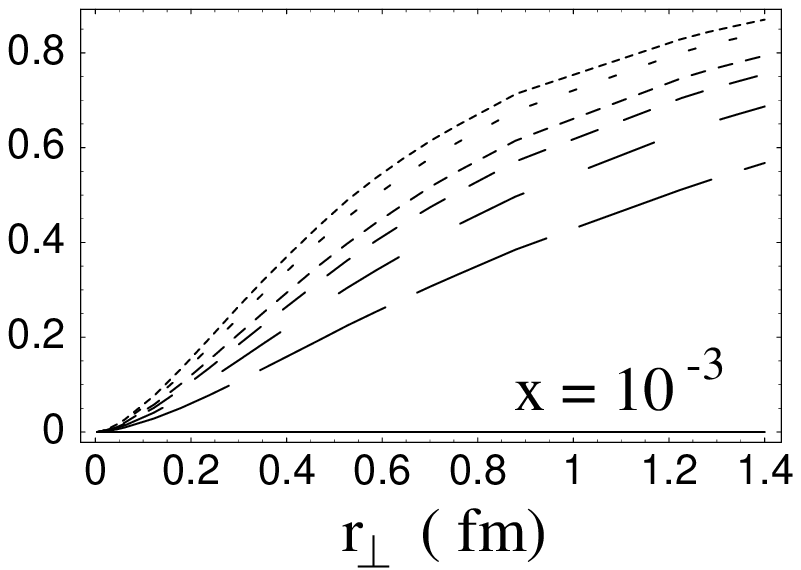,width=40mm, height=42mm} \\
\end{tabular}
  \caption[]{\it The function $\tilde N^D_A(x_0=10^{-2})$ is plotted versus transverse
  distance. The curves correspond to various nuclei (from up to down):
   Au, Nd, Mo, Zn, Ca, Ne. }
\label{solution}
\end{figure}

Fig.~\ref{solution} displays the solution $\tilde N^D_A$ as a
function of the transverse distance for several values of $x$ and
at fixed $x_0=10^{-2}$. The obtained curves reproduce the very
same pattern as in the case of a proton target \cite{DDLL}.  The
results for various nuclei can be used in order to study the $A$
dependence of the diffraction dissociation. In agreement with all
expectations, the unitarity bound $N^D_A=1$ is reached first by
the most heavy nucleus.
The dependence of the solutions  on the gap variable $x_0$ is very weak and quite similar to
the proton case of Ref. \cite{DDLL}.

\section{Saturation Scale}

Determination of the diffractive saturation scale $Q_s^D(A,x,x_0)$
from the solution $\tilde N^D_A$  is a subject of this section.
Following the same spirit of our previous works
\cite{LGLM,LL,me,DDLL} we  introduce several definitions of the
saturation scale while a variety of thus obtained  results will
indicate an uncertainty of the determined values.

For the step like function it is natural to define the saturation
scale as a position where $\tilde N_A^D=1/2$:
\begin{itemize}
\item {\bf Definition (a):}
\beq \label{defa}
\tilde N_A^D(R_s^D,x,x_0)\,=\,1/2\,,\,\,\,\,\,\,\,\,\,\,\, Q_s^D\,\equiv\, 2/R_s^D\,.
\eeq
\end{itemize}
The equality between the saturation radius $R_s^D$ and the
saturation momentum $Q_s^D$  is motivated by the double
logarithmic approximation. Though
 this approximation is formally  not justified, we still believe it to make reliable
estimates provided $Q_s^D$ is  large enough. The definition
(\ref{defa}) is analogous to the one proposed in Ref. \cite{LGLM},
$N(2/Q_s,x)=1/2$. If we recall that  $N^D_A=N^2_A$ at  $x=x_0$ and
postulate $Q_s^D(A,x_0,x_0) =Q_s(A,x_0)$ then we should require
\begin{itemize}
\item {\bf Definition (b):}
\beq \label{defb}
\tilde N^D_A(2/Q_s^D,x,x_0)\,=\,1/4\,.
\eeq
\end{itemize}
An alternative definition of the saturation scale could be one
motivated by the Glauber-Mueller formula \cite{DOF3} for
$N_A(r_{\perp},x)$ \beq \label{GM}
 N_A(r_\perp,x,x_0; b)\,=\, (1\,-\,e^{-\kappa_A(r_\perp,x,x_0)\, S_A(b)})\,\,,
\eeq with \beq \label{KAPPAGM} \kappa_A = \frac{3\,\pi^2\,\,
\alpha_S}{4}\,\,r^2_{\perp}\,A\,xG(x,\frac{4}{r^2_{\perp}})\,\,,
\eeq and $xG$ standing for the gluon density of a nucleon.

\begin{itemize}
\item {\bf Definition (c):}
\beq \label{defc}
\kappa^D_A(2/Q_s^D,x,x_0)\,=\,1/2\,.
\eeq
\end{itemize}

\begin{figure}[htbp]
\begin{tabular}{c c c c }
def (a) & def (b) & def (c) & def (d) \\
 \epsfig{file=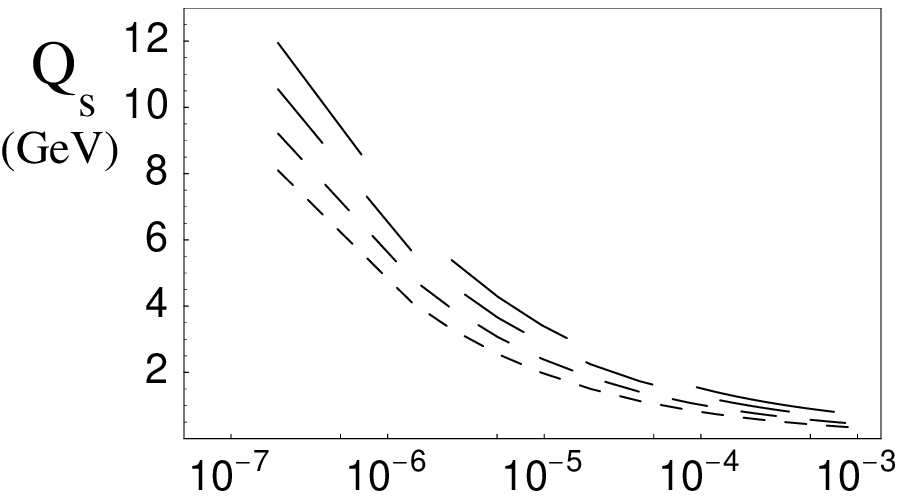,width=41mm, height=36mm}&
\epsfig{file=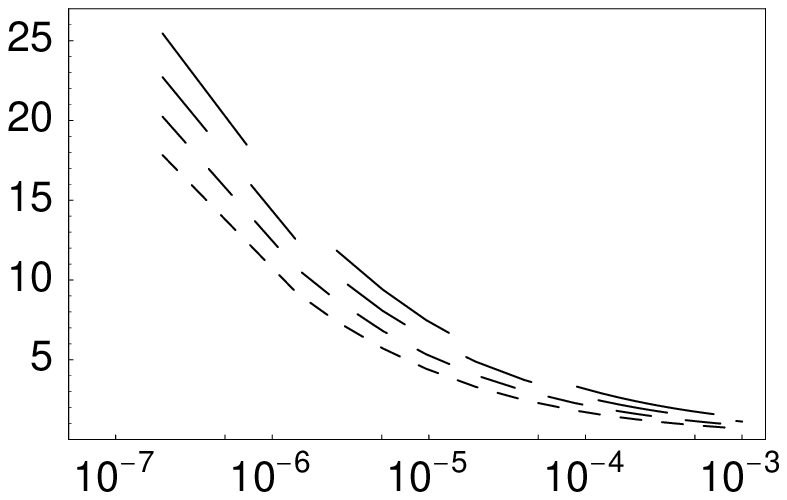,width=38mm, height=36mm}&
\epsfig{file=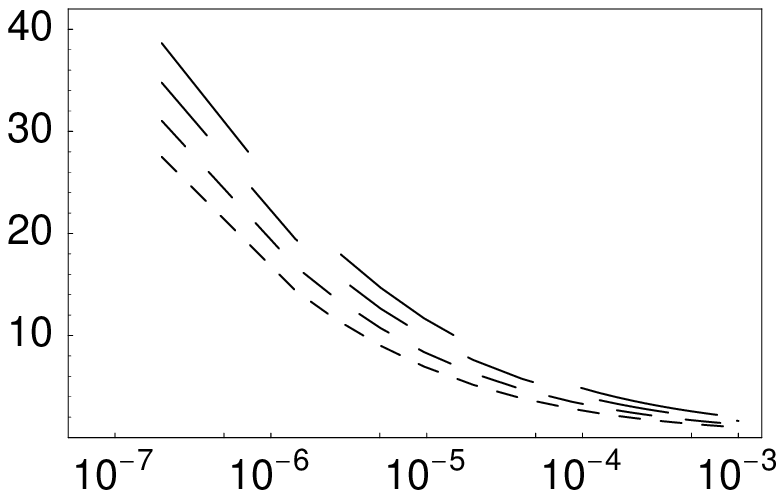,width=38mm, height=36mm}&
 \epsfig{file=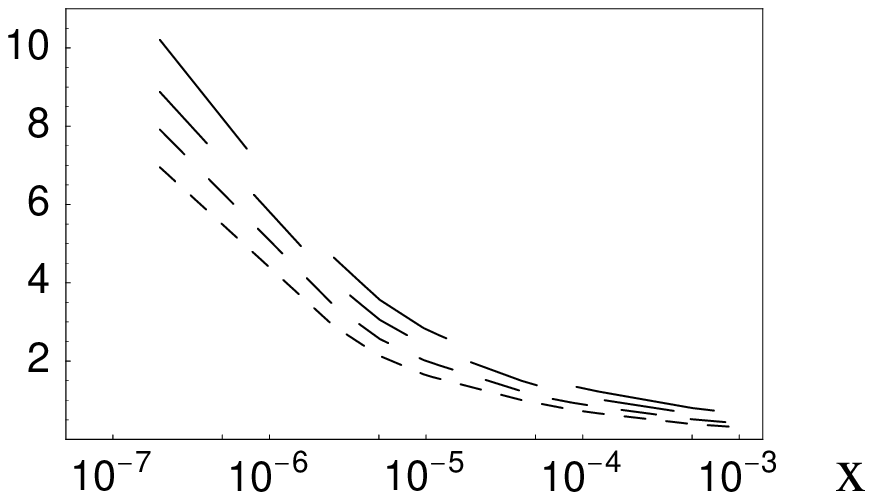,width=41mm, height=36mm} \\
\end{tabular}
  \caption[]{\it The saturation scale $Q_s^D(x_0=10^{-2})$ is plotted versus $x$.
The curves correspond to (from up to down) Au, Mo, Ca, and Ne.}
\label{Qs12}
\end{figure}

\begin{figure}[htbp]
\begin{tabular}{c c c c }
def (a) & def (b) & def (c) & def (d) \\
 \epsfig{file=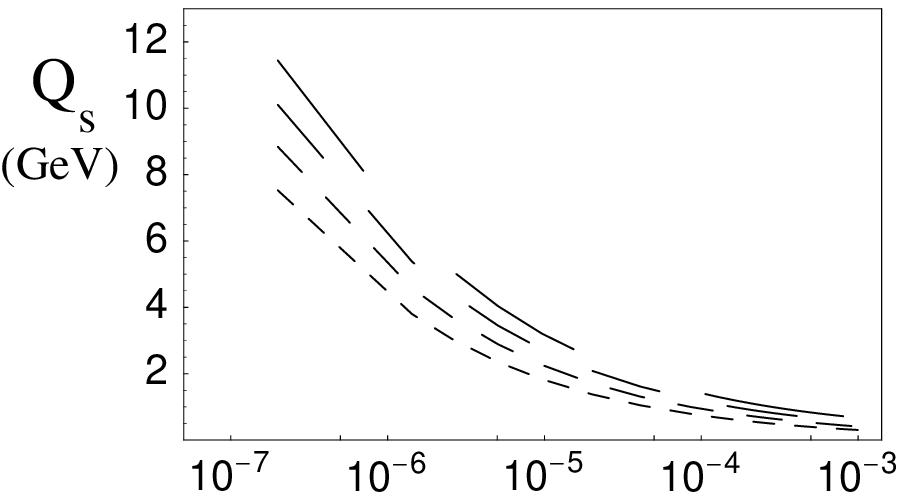,width=41mm, height=36mm}&
\epsfig{file=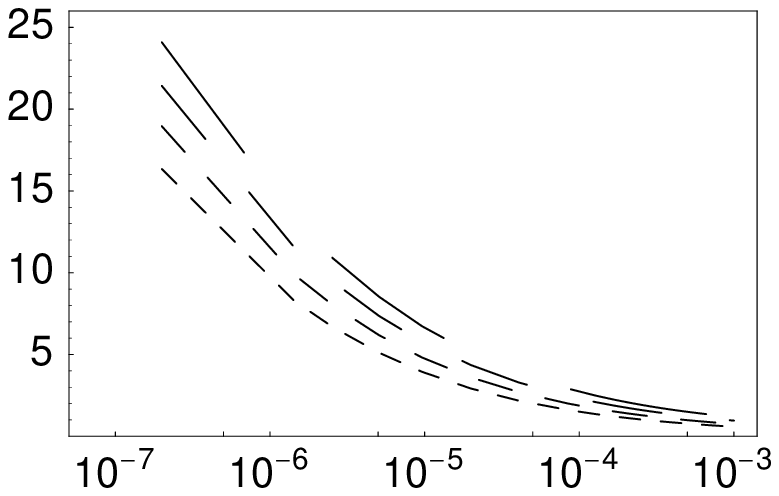,width=38mm, height=36mm}&
\epsfig{file=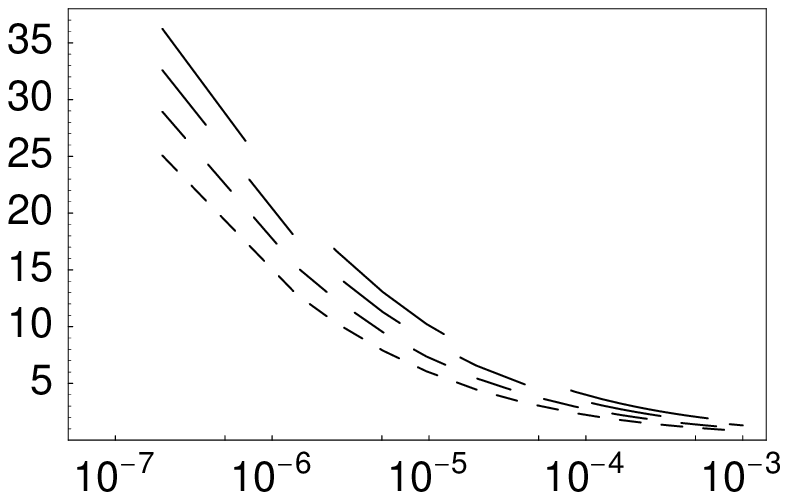,width=38mm, height=36mm}&
 \epsfig{file=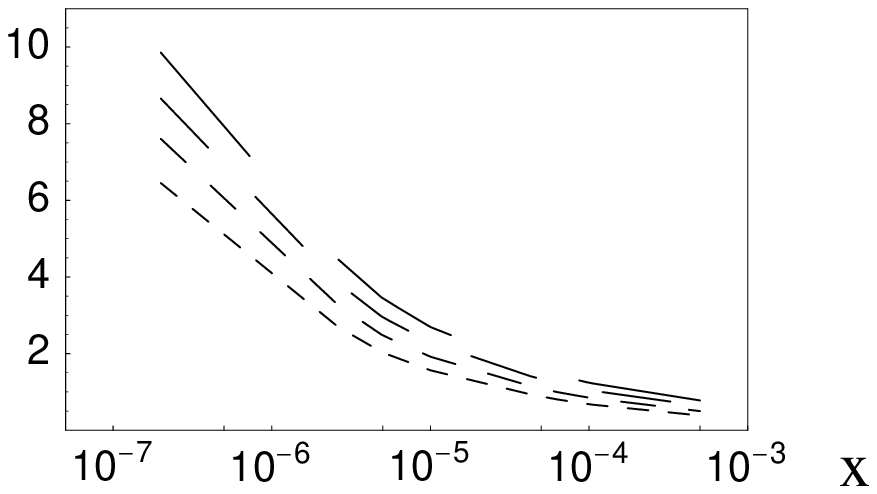,width=41mm, height=36mm} \\
\end{tabular}
  \caption[]{\it The saturation scale $Q_s^D(x_0=10^{-3})$ is plotted versus $x$.
The curves correspond to (from up to down) Au, Mo, Ca, and Ne.}
\label{Qs13}
\end{figure}

The obtained saturation scales are depicted in Fig. \ref{Qs12} for
$x_0=10^{-2}$ and in Fig. \ref{Qs13} for $x_0=10^{-3}$.
The saturation scales are practically $x_0$ independent. The
definition (d) is related to  scaling properties of the function
$N^D_A$ and will be discussed in the next section.

It is important to learn about  $x$ and $A$  dependencies of the
saturation scale. To this goal, we assume the following
parametrization: \beq\label{qsat}
Q_s^D(A,x,x_0)\,=\,Q_{s\,0}^D\,(x_0)\,A^p\,\,x^{-\lambda}\,. \eeq
In fact, the parametrization (\ref{qsat}) is a good approximation
for the obtained values of the saturation scales  with
$$
\lambda\,=\,0.37\,\pm\,0.04\,,$$ and $p$ given in Table \ref{tA}.
All the values in the table are central values given with errors
less than 10\%.

 The
$x_0$-dependence of the saturation scale is very weak. If we try
to use the power law parametrization $Q_{s\,0}^D\,\sim x_0^\beta$
then $ \beta\,=\,0.055\,\pm\,0.035\,$. The large error reflects a
significant $x$-dependence of $\beta$ as well as its sensitivity
to a saturation scale definition.

\begin{table}
\begin{minipage}{9.0 cm}
\center{
\begin{tabular}{||l||c|c|c|c|c||}
Nuclei $\,\,\backslash \,\,x$ & $10^{-7}$ & $10^{-6}$ & $10^{-5}$ & $10^{-4}$ & $10^{-3}$  \\
\hline \hline
 Light &                                       0.15    &           0.20  &    0.24   &     0.29     & 0.35        \\
\hline
 Heavy &                                    0.15     &       0.19   &    0.22      &     0.25    &   0.29     \\
\hline
 All   &                                        0.15     &       0.19   &    0.23     &     0.27    &   0.32    \\
\hline
\end{tabular}}
\end{minipage}
\begin{minipage}{7.0 cm}
\caption{The power $p(x)$ for various values of $x$.}
\label{tA}
\end{minipage}
\end{table}

It is important to stress that the obtained power $\lambda$
coincides with the corresponding power of the saturation scales
$Q_s$ \cite{me} and $Q_s^D$ \cite{DDLL}. The error obtained for $\lambda$ is purely
numerical. We have to admit that two independent collaborations \cite{num} found
$\lambda\simeq 2\,\as$ which exceeds our value by about 25\%.
This discrepancy may result from
different definitions of the saturation scale. We remind that in our approach we actually
compute the saturation radius and not a momentum. A transition to the latter is likely to be more
complicated compared to the relation $Q_s=2/R_s$. Yet, our estimates of the saturation scale
based on the solutions in  momentum space do not indicate any dramatic change in $\lambda$. In addition, we have want to emphisize that we use completely
diffrent initial conditions compared to Ref. \cite{num}. It was shown in
Ref. \cite{LTGS} that in fact solutions to  the non-linear
evolution equation crucially depend on a choice of initial conditions.

\section{Geometrical Scaling}

In  Ref. \cite{me} the function $\tilde N$ was shown to display a
phenomenon of geometrical  scaling while Ref. \cite{DDLL} presents
a detailed analysis of the scaling for diffractive dissociation
off proton. As can be expected, the function $\tilde N^D_A$
displays the very same property and in this section we give a
brief illustration of the phenomena. In the saturation region the
scaling   implies the amplitude to be a function of only one
variable $\tau= r_\perp^2\cdot (Q_s^D(A,x,x_0))^2$:
\beq\label{SCALING} \tilde N^D_A(r_\perp,x,x_0)\,=\,\tilde
N^D_A(\tau) \eeq

Let us define the following derivative functions while the second
equalities hold if the scaling behavior (\ref{SCALING}) is
assumed: \beq
 N_y^D(r_\perp,A,x,x_0)\,\equiv\,-\,\frac{\partial \tilde N^D_A}{\partial Y}\,=\,
  \frac{d \tilde N^D_A}{d \tau}\,\tau\,\frac{\partial\ln (Q_s^D)^2}{\partial \ln x}\,,
\label{DX}
\eeq
\beq
 N_r^D(r_\perp,A,x,x_0)\,\equiv\, r_\perp^2\,\frac{\partial \tilde N^D_A}{\partial r_\perp^2}\, =\,
 \frac{d \tilde N^D_A}{d \tau}\,\tau\,,
\label{DR}
\eeq
\beq
 \Re(r_\perp,A,x,x_0)\,\equiv\,-\frac{\partial \tilde N^D_A}{\partial Y_0}\,=\,
  \frac{d \tilde N^D_A}{d \tau}\,\tau\,\frac{\partial\ln (Q_s^D)^2}{\partial \ln x_0}\,.
\label{DX0} \eeq If the scaling behavior (\ref{SCALING}) takes
place indeed, then both  ratios $N_y^D/N_r^D$ and $\Re/N_r^D$ are
$r_\perp$ independent functions.

\begin{figure}[htbp]
\begin{tabular}{c c c c}
$Ne\,\,\,x=10^{-3}$ &$Ne\,\,\,x=10^{-4}$ &$Ne\,\,\,x=10^{-5}$ &$Ne\,\,\, x=10^{-6}$ \\
 \epsfig{file=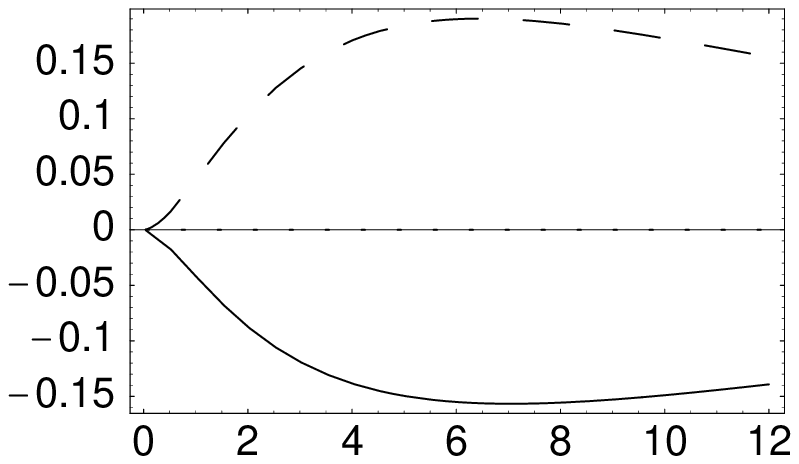,width=40mm, height=36mm}&
\epsfig{file=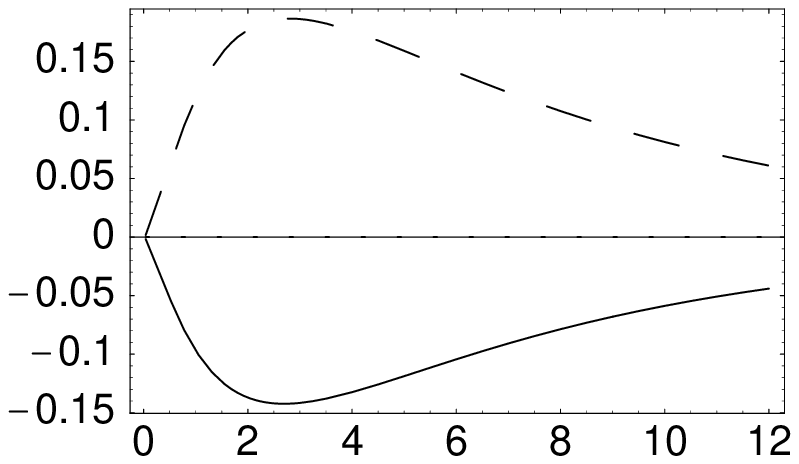,width=40mm, height=36mm}&
 \epsfig{file=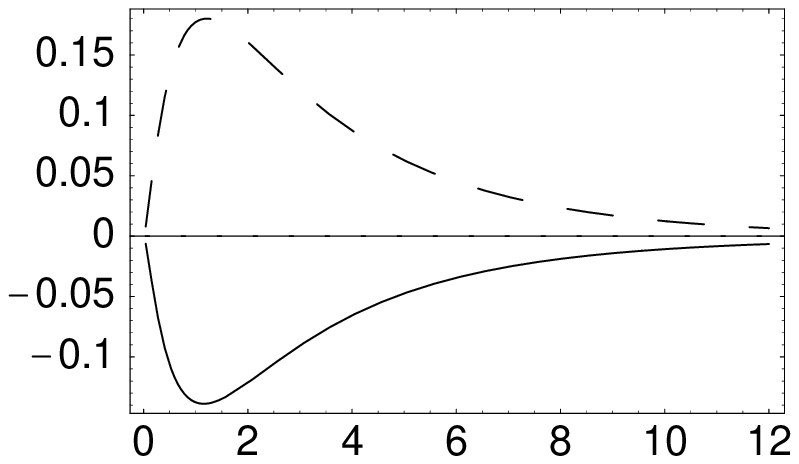,width=40mm, height=36mm}&
\epsfig{file=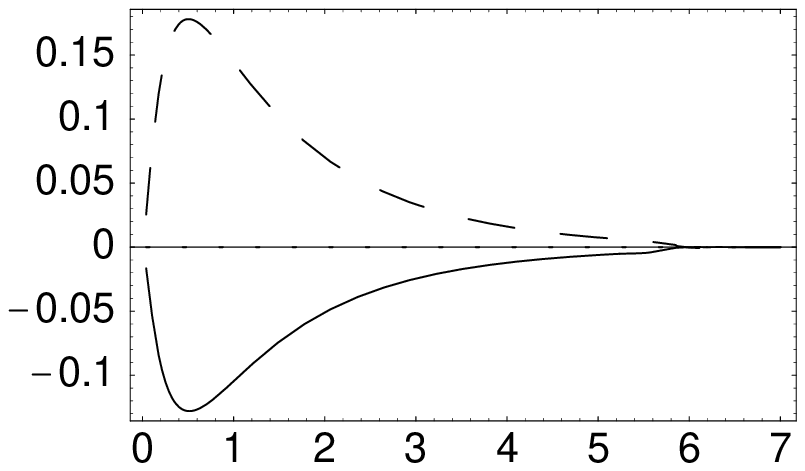,width=40mm, height=36mm}\\
$Mo\,\,\,x=10^{-3}$ &$Mo\,\,\,x=10^{-4}$ &$Mo\,\,\,x=10^{-5}$ &$Mo\,\,\, x=10^{-6}$ \\
 \epsfig{file=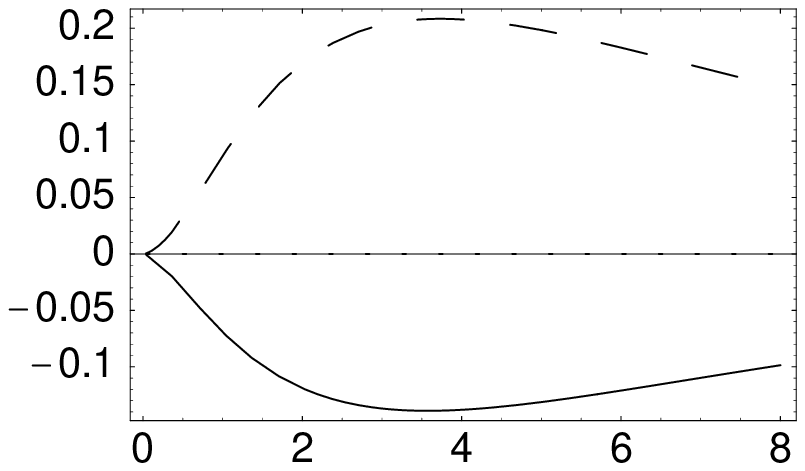,width=40mm, height=36mm}&
\epsfig{file=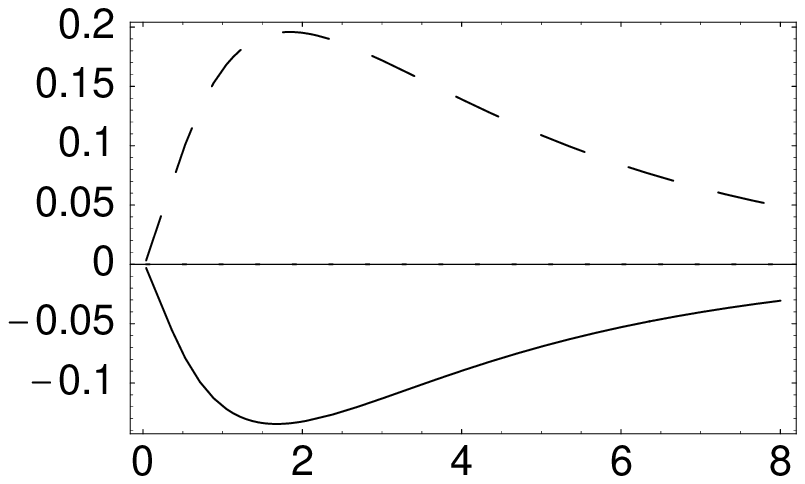,width=40mm, height=36mm}&
 \epsfig{file=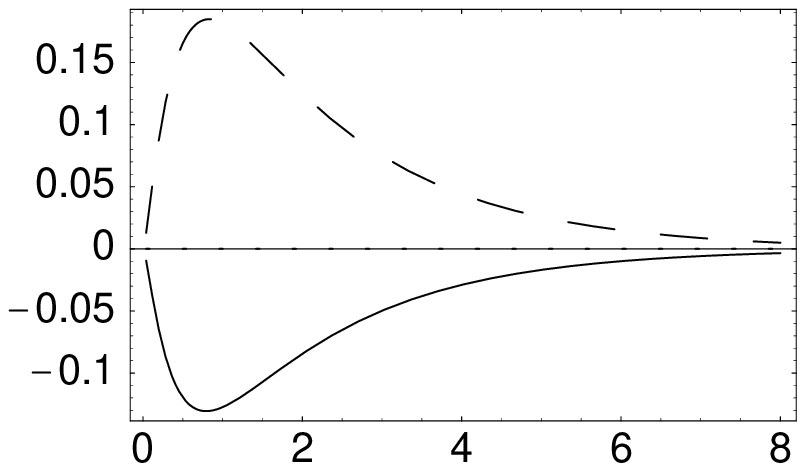,width=40mm, height=36mm}&
\epsfig{file=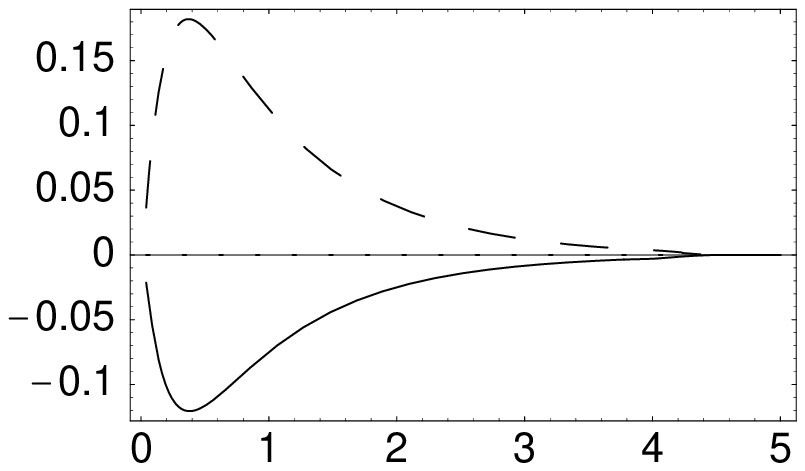,width=40mm, height=36mm}\\
$Au\,\,\,x=10^{-3}$ &$Au\,\,\,x=10^{-4}$ &$Au\,\,\,x=10^{-5}$ &$Au\,\,\, x=10^{-6}$ \\
 \epsfig{file=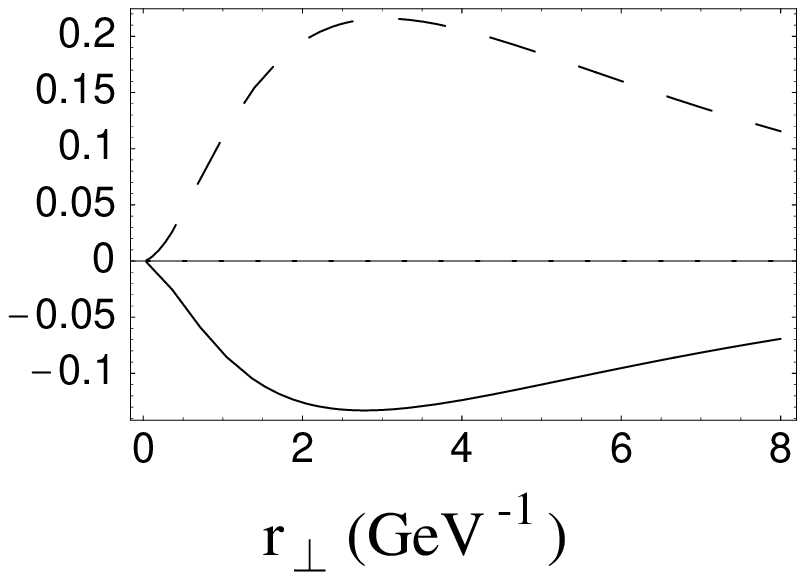,width=40mm, height=40mm}&
\epsfig{file=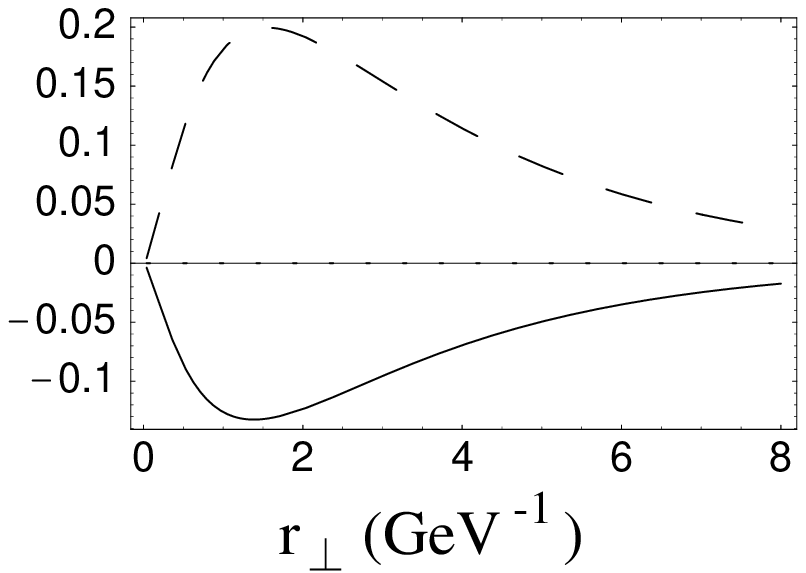,width=40mm, height=40mm}&
 \epsfig{file=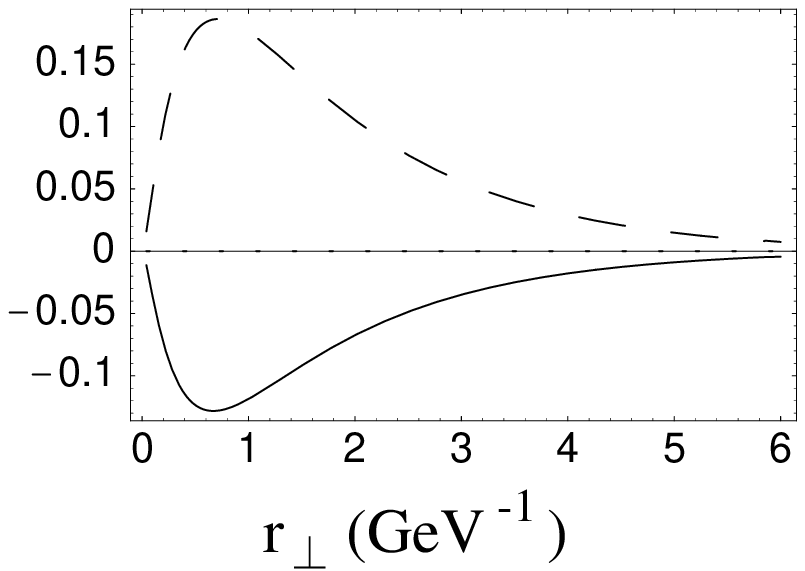,width=40mm, height=40mm}&
\epsfig{file=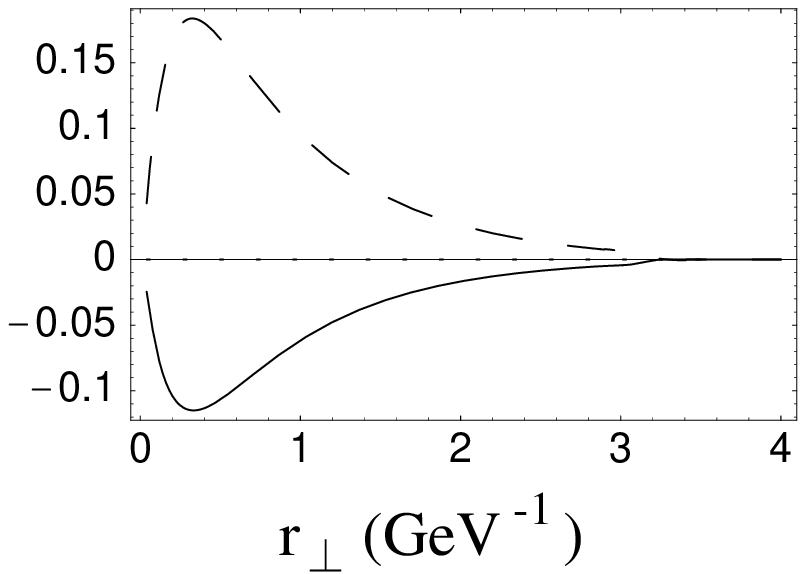,width=40mm, height=40mm}\\
\end{tabular}
  \caption[]{\it The derivative functions $N_r^D$ (dashed line) and $N_y^D$ (solid line)
as functions of transverse distance  at fixed $x_0=10^{-2}$. }
\label{scal_x}
\end{figure}

Let us first consider the scaling with respect to $x$. Fig.
\ref{scal_x} presents the derivatives
 $N_y^D$ and $N_r^D$  as functions of transverse distance $r_\perp$ at fixed $x_0=10^{-2}$.
Both functions  $N_y^D$ and $N_r^D$ have extremum placed at the
same distance depending on $x$ and atomic number $A$. This is a
consequence of the scaling behavior (\ref{SCALING}) and equations
(\ref{DX}) and (\ref{DR}). The extremum occures at certain
$\tau_{max}$, such that $\tilde N_A^{D\,\prime}
(\tau_{max})=-\tau_{max} \tilde
N_A^{D\,{\prime\prime}}(\tau_{max})$. In Fig. \ref{scal_x},
$\tau_{max}$ is approached by varying $r_\perp$ at fixed $x$.
Alternatively it can be reached by varying $x$  at fixed
$r_\perp$.

The position of the maximum $\tau_{max}$ can be used as  another
definition of  saturation scale:
\begin{itemize}
\item {\bf Definition (d):}
\beq \label{defd} \tau (r_\perp=2/Q_s^D)\,=\,\tau_{max}\,. \eeq
\end{itemize}
The saturation scale estimated from definition (d) is presented in
Figs. \ref{Qs12} and \ref{Qs13}.

For the sake of briefness we skip plots representing the ratios
$N_y^D/N_r^D$ and $\Re/N_r^D$. The $r_{\perp}$ independence of the
ratios is approximately reproduced. Thus the scaling with respect
to $x$ variable is established. Within relative error of order
20\% the resulting ratios do not depend neither on $x$ no on the
atomic number $A$. This observation is consistent with the power
law ansatz for the saturation scale (\ref{qsat}) and the
conclusions of the previous section. The results on the scaling
practically do not alter when $x_0$ is varied.

\section{$\sigma^{diff}/\sigma^{tot}$}

In this section we consider a ratio of the inclusive diffractive
dissociation to the total inclusive production. The mass maximally
produced in the diffractive process can be related to  the minimal
rapidity gap $Y_0=\ln 1/x_0$:
$$M^2\,=\,Q^2\,(x_0/x\,-\,1)\,.$$
At $x=x_0$ diffraction reduces to pure elastic scattering. We set
$x_0=10^{-2}$ as a minimal gap allowed in our calculations.

\begin{figure}[htbp]
\begin{tabular}{c c c}
$Q^2=1\,GeV^2$  & $Q^2=10\,GeV^2$ & $Q^2=80\,GeV^2$   \\
\epsfig{file=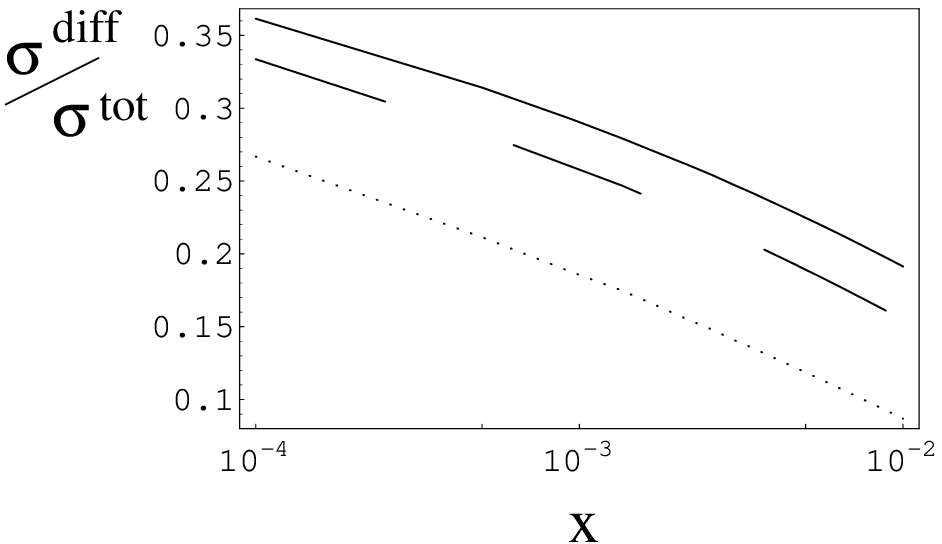,width=60mm, height=42mm}&
\epsfig{file=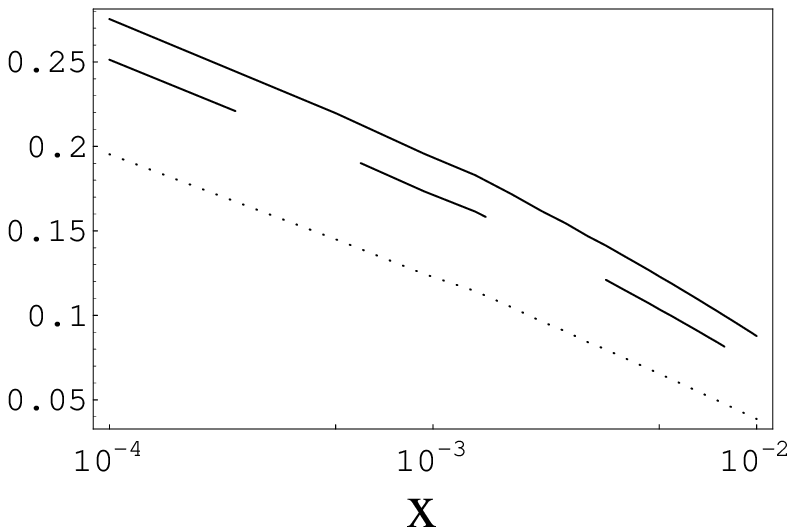,width=52mm, height=42mm}&
 \epsfig{file=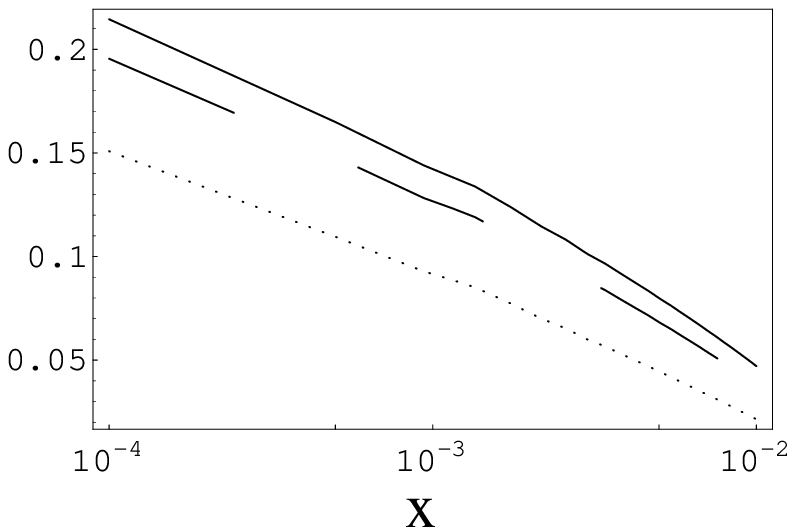,width=52mm, height=42mm}  \\
\end{tabular}
  \caption[]{\it The ratio of inclusive diffractive dissociation to total inclusive
  production  is plotted versus $x$.
The curves correspond to Au (solid line), Mo (dashed line),  and
Ne (dotted line) nuclei.} \label{ratplot}
\end{figure}

Fig. \ref{ratplot} displays the ratio $\sigma^{diff}/\sigma^{tot}$
as a function of $x$ for fixed values of the photon virtuality
$Q^2$. This ratio measures the value of shadowing corrections and
it grows as $x$ decreases tending to the unitarity bound 1/2. For
heavier nuclei,  diffraction dissociation is larger as a result of
 stronger nuclear shadowing.

\begin{figure}[htbp]
\begin{minipage}{9.0cm}
\epsfig{file=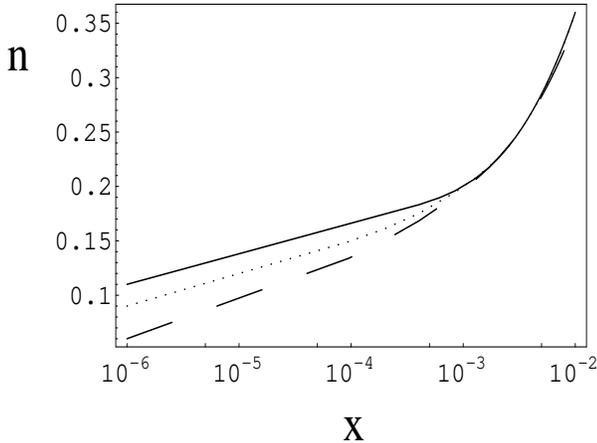,width=80mm, height=60mm}
\end{minipage}
\begin{minipage}{7.0 cm}
\caption{ \it The exponent $n$ is plotted versus $x$ for several
values of $Q^2$: $Q^2=80\,GeV^2$ (solid line), $Q^2=10\,GeV^2$
(dotted line), and $Q^2=1\,GeV^2$ (dashed line).} \label{adep}
\end{minipage}
\end{figure}

It is worth  to investigate the $A$-dependence of
$\sigma^{diff}/\sigma^{tot}$ which in perturbative QCD is
proportional to $A^{1/3}$ (times $\sigma^{diff}/\sigma^{tot}$ for
a proton). We assume the ratio $\sigma^{diff}/\sigma^{tot}$ to
have the power law dependence on $A$:
$$
\sigma^{diff}/\sigma^{tot}\,\,\sim\,\,A^{n(x)}\,.
$$
It is clear that in deep saturation regime $n$ should vanish
leading to the $A$-independent behavior
($\sigma^{diff}/\sigma^{tot}=1/2$). Fig. \ref{adep} displays the
function $n(x)$ for all nuclei in consideration.

It is important to note that the results obtained are in agreement
with the results of Ref. \cite{GLLMT} where the ratio
$\sigma^{diff}/\sigma^{tot}$ was considered on a basis of the
Glauber-Mueller formula.

\section{Theoretical  discussion.}

One of the most surprising results of our calculations is the
geometrical scaling behavior which holds even at sufficiently
short distances (large values of photon virtualities $Q^2$). At
short distances the  imaginary part of the dipole elastic
amplitude has the form
$$
N(A;r_{\perp},x)\,\propto
\,r^2_{\perp}\,xG_A(x,\frac{4}{r^2_{\perp}})/\pi R^2_A
\,\approx\,\,A^{\frac{1}{3}}\,r^2_{\perp}\,xG_{proton}(x,\frac{4}{r^2_{\perp}}).
$$

The scaling means that this function is a function of only one
variable  $\tau = r^2_{\perp}\,Q^2_s(A;x)$. Such a nontrivial
behavior should  certainly  result from a kind of interplay
between three variables $r_{\perp}$, $x$ and $A$. For a proton
target it was shown in Ref. \cite{MLGS} that the scaling holds in
a wide range of distances even for large virtualities $Q^2 \approx
4/r^2_{\perp} \,>\,Q^2_s(x)$:
 \beq \label{SCCON}
Q^2_s\,\,\leq\,\,Q^2\,\,\leq \,Q^4_s/\Lambda^2 \,.\eeq

However, for the geometrical scaling to hold in the case of
nuclear targets,  an additional condition on the value of the
saturation scale is required. Namely, the condition $Q^2_s(A;x)
\,\gg\,\Lambda^2\,\, A^{\frac{1}{3}}$ should be fulfilled in order
to justify the interplay between $A$ and the rest of the
variables. As was shown in Ref. \cite{LTGS}, in the region of low
$x$ and large photon virtualities the dipole amplitude looks as
follows

\beq \label{DLAA} N(A;\xi = \ln(1/(r^2_{\perp}\,\Lambda^2)), y -
y_0 = \ln(x_0/x))\,\,\approx\,\,e^{2\,\sqrt{
\bar{\alpha}_s\,\xi\,(y - y_0)}\,\,+\,\,\xi_A\,-\,\xi}\,\,, \eeq
where $\xi_A = \frac{1}{3} \ln A$.

 Saturation scale can be found from the condition
$N \,=\, Const (A,r_{\perp},x)$, which leads to the equation for
the saturation scale $\xi_s \,=\,\ln(Q^2_s/\Lambda^2)$:

\beq \label{SATSCA} 4\,\,\bar{\alpha}_s\,\xi_{sat}\,(y -
y_0)\,=\,(\xi_{sat} \,-\,\xi_A)^2\,\,. \eeq

As a result,  $\xi_{sat}(y = y_0) = \,\xi_A$ while $\xi_{sat}
\,=\, 4\,\,\bar{\alpha}_s (y - y_0)\,+\,2\,\xi_A$ for $\xi_{sat}
\gg \xi_A$. Let us expand \eq{DLAA} in the vicinity of $Q^2
\approx Q^2_s$. Introducing $\xi = \Delta \xi + \xi_{sat}$ with
$\Delta \xi\,=\,\ln(Q^2/Q^2_s)$ and considering $\Delta \xi
\,\ll\,\xi_{sat} $ we get

\beq \label{EXPA} 2 \,\sqrt{ \bar{\alpha}_s\,\xi\,(y - y_0)}
\,+\,\xi_A\,-\,\xi \,=\,-\, \frac{1}{2}\,\Delta \xi \,+\,\Delta
\xi \,(\frac{\xi_A}{\xi_{sat}}) \,+\,O\left( (\frac{\Delta
\xi}{\xi_{sat}})^2\right)\,. \eeq

We are interested in the region where an approximate geomentrical scaling behavior
can be seen. So,
we consider $\Delta \xi \approx 1$ and
for the scaling
behavior to hold we have to  assume that
$$ \Delta \xi\,\ll\,\xi_{sat}\,\,\,\mbox{and}\,\,\xi_A \,\ll\,\xi_{sat}\,.$$
The first inequality leads to $Q^2 \,<\,Q^4_s/\Lambda^2$ while the
second one gives $Q^2_s > \Lambda^2 \,A^{\frac{1}{3}}$.

In the particular model given by Eq. (\ref{DLAA}) and at low $x$,
the $A$ dependence of the saturation scale squared is
$Q^2_s\,\propto\,A^{\frac{2}{3}}$. For such a dependence the
condition $Q^2_s > \Lambda^2 \,A^{\frac{1}{3}}$ means that $x$ is
supposed to be sufficiently small. Our numerical solution,
however,  shows rather different dependence on $A$
($Q^2_s\,\propto\,A^{\frac{1}{3}}$) and the requirement $Q^2_s
>\,\Lambda^2 \,A^{\frac{1}{3}}$ has   numerical justification
only.  We would like to recall that the above model is correct at
very large photon virtualities only. A natural question arises how
to extend this model to smaller virtualities in the range defined
by \eq{SCCON}. We will discuss this issue below.

The above discussed simple model allowed us to illustrate that
the theoretical expectations for the $A$-dependence of the
saturation scale are quite different from the numerical
calculations. A possible reason for such different $A$-dependence
is in the fact that  $x = 10^{-2} - 10^{-5}$ are not small enough
to apply this model. For rather low energies the saturation scale
squared obtained from  \eq{SATSCA} is proportional to
$A^{\frac{1}{3}}$ in agreement with the numerical calculations. In
fact, we are still far away from the low $x$ region  where the
collective effects related to saturation  are strong. This can be
seen from the ratio $\sigma^{diff}/\sigma^{tot}$ which is not
close to the saturation limit $1/2$ and moreover exhibits some $A$
dependence.

\begin{figure}[htbp]
\begin{center}
\epsfig{file=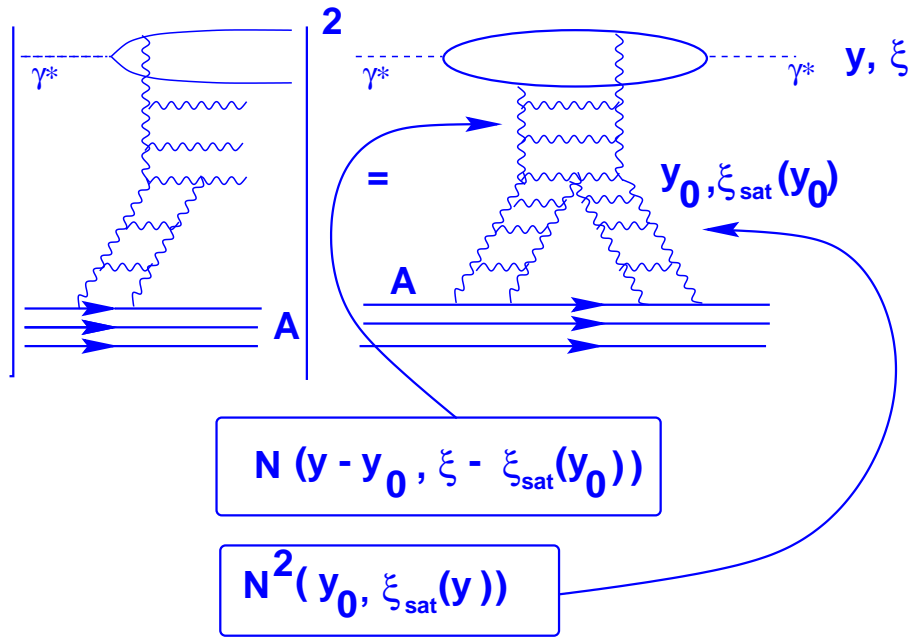,width=120mm}
\end{center}
  \caption[]{\it The diffractive production of large mass (M)  ($y - y_0
=
\ln M^2$) in DIS.}
 \label{disdifsat}
\end{figure}

Now, let us try to analyze the diffraction dissociation in DIS
using the same approach as in the discussed model. We consider so
large  initial virtualities that we can restrict ourselves to
solutions of the linear DGLAP evolution equation. The diagram
which contributes to high mass diffraction is shown in Fig.
\ref{disdifsat}. It is well known (see Ref.\cite{LW} for example)
that this diagram can be expressed in the form

\beq \label{ND} N^D(y - y_0 =\ln M^2, \xi)\,=\,\int d \,\xi' N(y -
y_0, \xi - \xi')\,\,N^2(y_0, \xi')\,\, \eeq

with $N(y - y_0, \xi - \xi')$ given for $\xi - \xi'\gg \xi_{sat}$
by

\beq \label{N}
  N(y - y_0, \xi - \xi') \,\,=\,\,\int \frac{d f}{2\pi
i}\,\,e^{\frac{\bar{\alpha}_s}{f}\,\,+\,(f - 1)\,(\xi -
\xi')}\,\,\propto\,e^{ 2 \sqrt{\bar{\alpha}_s\,(y - y_0)\,(\xi -
\xi')}\,\,- \xi\,+\,\xi'}\,\,. \eeq

It can be seen that Eq. (\ref{ND}) is a solution to Eq.
(\ref{DDEQ}) provided  the non-linear terms are neglected in this
equation. Moreover, \eq{ND} satisfies the correct initial
conditions. The non-linear corrections are neglected since we wish
to estimate the saturation scale which corresponds to sufficiently
small $N^D$ (in semiclassical approach, for example, $N^D
\,\propto\,\as$). It is important to emphasize that the $A$
dependence in \eq{ND} comes from the initial conditions only.

Assume the value of $x_0$ to be so small that $N(y_0, \xi') $ is
saturated at the saturation scale $\xi' = \xi_{sat}(y_0)$. In this
case, the typical value of $\xi'$ in the integrand of \eq{ND} is
$\xi' \simeq \xi_{sat}(y_0)$. Indeed, for $\xi' \,<\,
\xi_{sat}(y_0)$, $N(y_0, \xi') \approx const(y_0,\xi')$ and the
main contribution to the integral comes from the upper limit of
the integration $\xi' \,=\, \xi_{sat}(y_0)$. In the region $\xi'
\,>\, \xi_{sat}(y_0)$ the integrant falls down exponentially as a
function of $\xi'$. As a result, the diffractive amplitude $N^D$
is approximately equal

\beq \label{NDSAT}
N^D(y - y_0 =\ln M^2, \xi)\,=\, N(y - y_0, \xi -
\xi_{sat} (y_0))\,\,N^2(y_0, \xi_{sat})\,\, .
\eeq

Since $N(y_0, \xi_{sat})\,\,=\,\, Const(y_0)$ we have \beq
\label{NDSAT1} N^D(y - y_0 =\ln M^2, \xi)\,\propto\,e^{ 2
\sqrt{\bar{\alpha}_s\,(y - y_0)\,(\xi -\xi_{sat}(y_0))}\,\,-
\xi\,+\,\xi_{sat}(y_0)}\,\,. \eeq

\eq{NDSAT1} leads to the saturation scale given by the equation
\beq \label{SATSCD} 4\, \bar{\alpha}_s (y - y_0)\,= \xi^D_{sat}(y)
- \xi_{sat}(y_0)\,. \eeq

Recall that $\xi_{sat}(y_0) = 4\bar{\alpha}_s \,\,y_0  + \xi_A\,$
($ \xi_{sat}(y_0) \approx \xi_A$). Finally we obtain from
\eq{SATSCD} that \beq \label{SATDF} \xi^D_{sat}(y) = 4\,
\bar{\alpha}_s\,\,y \,\,+\,\, \xi_A\,. \eeq

It can be seen that this simple approach reproduces our numerical
result that the saturation scale for the diffractive production in
DIS has the very same $x$ dependence as the saturation scale for
the total DIS cross section. However, it is important to note that
the energy dependence of the saturation scale obtained from these
simple theoretical estimates turns out to be quite different from
the numerical results. Within the approximations made,
$\xi^D_{sat}(y)$ does not depend on $y_0$ ($x_0$) in accordance
with the numerical calculations.

Comparing Eqs. (\ref{SATSCA}) and (\ref{SATSCD}), the
$A$-dependence of the saturation scale in total and diffractive
productions appear to be quite different. Technically, this
difference arises from the extra $\xi'$ integration in \eq{ND} in
comparison with \eq{DLAA}. Keeping this in mind we are going to
reconsider saturation in total cross section. To this goal, we can
 rewrite the dipole amplitude $N$ as an integral over the intermediate
virtuality $\xi'$. As follows from the general properties of the
DGLAP equation this can be  always  done.

\begin{figure}[htbp]
\begin{center}
\epsfig{file=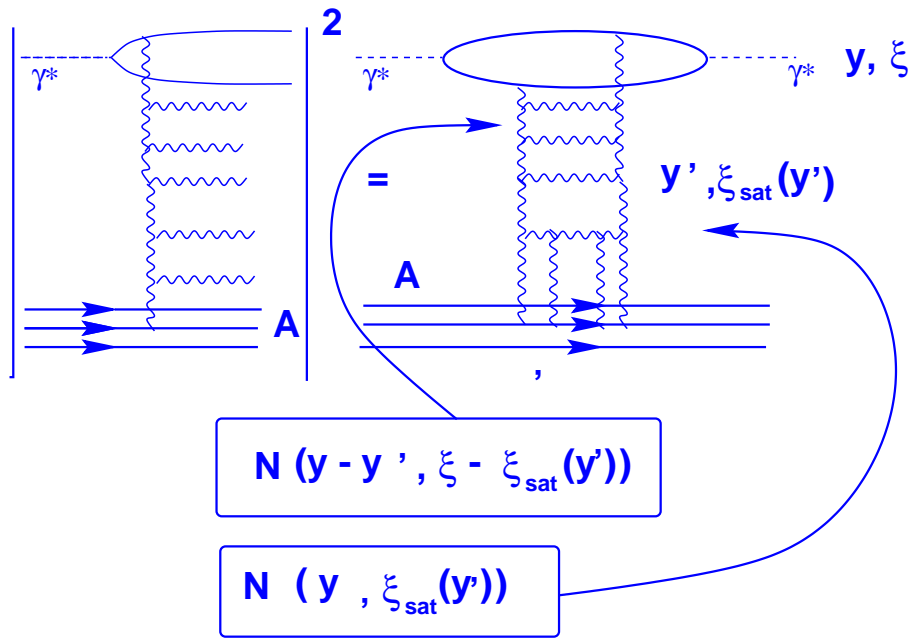,width=120mm}
\end{center}
  \caption[]{\it The total cross section for DIS with
initial conditions from the McLerran-Venugopalan model.}
 \label{distotsat}
\end{figure}

Fig. \ref{distotsat} displays the DIS process in which the initial
condition for the DGLAP linear evolution  is fixed by the
McLerran-Venugopalan model (see Ref. \cite{HDQCD}). In other
words, it is assumed that at sufficiently low energies ($y'$ in
Fig. \ref{distotsat}),  saturation at $Q^2_s(A, y')
\,\gg\,\,\Lambda^2$ is reached due to strong gluonic fields in  a
nuclear target. In this case the integral over $\xi'$  is
dominated by $\xi' \simeq \xi_{sat}(y')$ and the the dipole
amplitude $N$ is given by \beq \label{NML} N(A;\xi =
\ln(1/(r^2_{\perp}\,\Lambda^2), y - y' =
\ln(x'/x))\,\,\approx\,\,e^{2\,\sqrt{
\bar{\alpha}_s\,(\xi\,-\,\xi_{sat}(y') (y -
y')}\,\,+\,\,\xi_{sat}(y')\,-\,\xi}\,\,. \eeq Since $\xi_{sat}(y')
\,\approx\,\xi_A$ in the initial condition for the saturation
scale,
  the following expression  is obtained:
\beq \label{SCTOT} \xi_{sat}(y)\, = \,4\,\bar{\alpha}_s\,(y -
y')\,\,+\,\,\xi_{sat}(y') = 4\,\bar{\alpha}_s\,(y -
y')\,\,+\,\,\xi_A \,\, . \eeq

 The saturation scale defined in Eq. (\ref{SCTOT}) is proportional to
$A^{\frac{1}{3}}$ in a contrast to Eq. (\ref{SATSCA}). A  natural
question arises: what is wrong with the first approach? \eq{DLAA}
satisfies all correct initial conditions at short distances. In
the integration over $\xi'$ the main contribution comes from the
upper limit of the integration $\xi' \approx \xi_{sat}$. The
integral over $\xi'$ in the saturation region looks as follows:
$$\int^{\xi_{sat}}\,d \xi' \,e^{\xi'}\,\,\rightarrow \,\,e^{\xi_{sat}}.$$

For $\xi' \,>\,\xi_{sat}$ the integral is $\int_{\xi_{sat}}\,d
\xi'\,\,e^{-\xi'}$. This simple form of the integral is valid in
the region  $Q^2 \,>\,Q^4_s/\Lambda^2$. This contribution is small
but the estimates in the beginning of this section show that at
sufficiently short distances defined by \eq{SATSCA}  even such a
small contribution  may reach so large values that we have a dense
system of partons.

Actually, \eq{DLAA} is \beq \label{DLAUN} N(A;\xi =
\ln(1/(r^2_{\perp}\,\Lambda^2), y - y_0 = \ln(x_0/x))\,\,=
r^2_{\perp}\,A\,xG(x,\frac{4}{r^2_{\perp}})/\pi R^2_A \eeq in
usual notations  with $xG$ being in double log approximation of
pQCD. \eq{DLAUN} is correct at least at very short distances.

Within the same notations \eq{NML}  looks differently \beq
\label{NMLUN} N(A;\xi = \ln(1/(r^2_{\perp}\,\Lambda^2)), y - y_0 =
\ln(x_0/x))\,\,=
r^2_{\perp}\,A\,xG(x,\frac{4}{r^2_{\perp}\,Q^2_s(A;y')})/\pi
R^2_A\,.
 \eeq

As we  discussed (see Ref. \cite{MLGS} for more details) the
geometrical scaling behavior of \eq{NMLUN} is preserved till
$r^2_{\perp}\,\approx 4\Lambda^2/(A^{\frac{1}{3}} \,Q^4_s(y'))$.
For shorter distances we expect  the regime discussed in the
beginning of this section to take place with  \eq{DLAUN} being
correct. This should happen at $ Q^2 >
 Q^4_{sat} /(A^{\frac{1}{3}}\Lambda^2)$ and at $x\rightarrow 0 $. Unfortunately
all experimentally accecible values of $x$ are not small enough for \eq{DLAA}
to be seen in a nearest future.

To evaluate a typical scale of distances involved we recall that
 the first RHIC data are likely to reveal
$Q^2_s(\mbox{Gold},x=10^{-2}) \,\,\approx\,\,2\,\,GeV^2$
\cite{KLN}. Therefore, the  regime with $Q^2_s \,\propto
\,A^{\frac{2}{3}}$ is expected to start at $r^2_{\perp}
\,<\,0.01\,GeV^{-2} = 4\,10^{-4} fm^2$. As can be seen in Fig.
\ref{solution}, at so short distances and at  $ x = 10^{-2} -
10^{-6}$ the diffraction amplitude is far away from the saturation
region.

\section{Conclusions}

The non-linear evolution equation (\ref{DDEQ}) is solved
numerically for six real nuclear targets $Ne_{20}$, $Ca_{40}$,
$Zn_{70}$, $Mo_{100}$, $Nd_{150}$, and $Au_{197}$. The obtained
solutions
 display the very same pattern as in the case of proton
target \cite{DDLL}. These solutions are used to study the
$A$-dependence of single diffractive dissociation.

The saturation scale $Q_{s,\,A}^D$ is estimated. The fit to the
parameterization $Q_s^D(A,x,x_0)\,\sim\,A^p\,x^{-\lambda}\,$
reveals powers $p$ and $\lambda$. The results on $\lambda$
coincide with the corresponding power obtained for the proton case
\cite{DDLL} while the $A$-dependence is the same as found for the
saturation scale $Q_{s,\,A}$ - saturation scale obtained in total
production \cite{LL}. The function  $Q_{s,\,A}^D$ is found to be
almost independent on $x_0$.

The geometrical scaling  with respect to $x$ is well established
for all nuclei considered. The scaling holds within a few percent
accuracy and in the whole kinematic region investigated. As a
consequence, inclusive diffractive production off nuclei is
predicted to possess the scaling both with respect to $x$ and $A$.
Like for a proton target the scaling phenomena with respect to
$x_0$ set in at $x\ll x_0$ but is violated at $x\sim x_0$.

The $A$ and $x_0$ dependences of the determined saturation scale
can be given a theoretical explanation. However, the numerically
found $x$ dependence of the scale is weaker compared to the
theoretical estimates.

The ratio  $\sigma^{diff}/\sigma^{tot}$ was examined on a basis of
the solutions obtained. At $x\simeq 10^{-4}$ a significant
shadowing is expected of the order 25\% and it is larger for heavy
nuclei. The fact that this ratio turns out to be very close to the
estimates based on the Glauber-Mueller formula \cite{GLLMT} shows
that  the latter can be used as a simple approach for first
estimates of possible collective effects in the saturation region
of high parton density QCD.

{\bf Acnowledgements:} The authors wish to thank the DESY and
Hamburg University Theory Groups for their hospitality and
creative atmosphere during several stages of this work.
 Our special thanks go to Asher Gotsman,
 Uri Maor and Eran Naftali for very stimulating discussion on the subject
of this paper.

Part of this work done by M.L. was performed in the Technion. M.L.
is very grateful to Physics Department of the Technion and
especially to the members of the High Energy Group for warmness
and creative atmosphere.

This research was supported in part by the BSF grant $\#$ 9800276,
by the GIF grant $\#$ I-620-22.14/1999
  and by
Israeli Science Foundation, founded by the Israeli Academy of
Science and Humanities.

\end{document}